\documentclass{emulateapj}

\newcommand\etal{{\it et al. }}
\newcommand\kms{km~s$^{-1}$~}
\newcommand\msun{$M_\odot$}

\def\be{\begin{equation}}
\def\ee{\end{equation}}

\shorttitle{ALFALFA Minihalo Candidates}

\begin{document}
\title{Are Newly Discovered HI High Velocity Clouds Minihalos in the Local Group?}

\author {Riccardo Giovanelli\altaffilmark{1,2}, Martha P. Haynes\altaffilmark{1,2}, 
Brian R. Kent\altaffilmark{3}, Elizabeth K. Adams\altaffilmark{1}}


\altaffiltext{1}{Center for Radiophysics and Space Research, Space Sciences Bldg.,
Cornell University, Ithaca, NY 14853. {\it e--mail:} riccardo@astro.cornell.edu,
haynes@astro.cornell.edu, betsey@astro.cornell.edu}

\altaffiltext{2}{National Astronomy \& Ionosphere Center, Cornell University,
Space Sciences Bldg.,
Ithaca, NY 14853. The National Astronomy \& Ionosphere Center is operated
by Cornell University under a cooperative agreement with the National Science
Foundation.}

\altaffiltext{3}{Jansky Fellow, National Radio Astronomy Observatory, 520 Edgemont Rd., 
Charlottesville, VA 22903. The National Radio Astronomy Observatory is operated
by Associated Universities, Inc. under a cooperative agreement with the National Science
Foundation. {\it e--mail:}bkent@nrao.edu }

\begin{abstract}
A set of HI sources extracted from the north Galactic polar region 
by the ongoing ALFALFA survey has properties that are consistent with 
the interpretation that they are associated with isolated minihalos 
in the outskirts of the Local Group (LG). Unlike objects
detected by previous surveys, such as the Compact High Velocity Clouds
of Braun \& Burton (1999), the HI clouds found by ALFALFA do
not violate any structural requirements or halo scaling laws of the
$\Lambda$CDM structure paradigm, nor would they have been detected
by extant HI surveys of nearby galaxy groups other than the LG.
At a distance of $d$ Mpc, their HI masses range between 
$5\times 10^4d^2$ and $10^6d^2$ \msun ~and their HI radii between $<0.4d$ and
$1.6d$ kpc. If they are parts of gravitationally bound halos, the total masses 
would be on order of $10^8$--$10^9$ \msun, their 
baryonic content would be signifcantly smaller than the cosmic fraction of
0.16 and present in a ionized gas phase of mass well exceeding that of the neutral
phase. This study does not however prove that the minihalo interpretation is
unique. Among possible alternatives would be that the clouds are shreds of the Leading 
Arm of the Magellanic Stream.
\end{abstract}


\section {Introduction}\label{intro}

The $\Lambda$CDM paradigm which describes the evolution of structure predicts 
the existence of large numbers of low mass ($\lesssim 10^9$ \msun) halos. A 
cosmic census of dwarf galaxies at $z=0$ indicates that such objects are 
rarer than expected from numerical simulations. In different guises, this 
circumstance has been referred to as the ``the void phenomenon'' 
(Peebles 2001) or the ``missing satellite'' problem (Klypin \etal 1999; 
Moore \etal 1999). The gap between the number of Milky Way (MW) satellites 
expected from numerical simulations and that obtained from observations has 
progressively narrowed in recent years with the discovery of fossil stellar 
structures in optical wide--field surveys. While this technique has 
detected numerous dwarf satellites of the MW, no gas--rich systems have been
found and the number of known halos with circular velocities of 
$v_{circ}\lesssim 20$ \kms located in the general field remains very low. 
As is the case with the luminosity function of
optical galaxies, the mass function of extragalactic HI sources also has a
faint end slope much shallower than predicted by the Press--Schechter
formalism (Zwaan \etal l2006; Martin \etal 2009, in preparation). HI surveys are thus not
detecting enough low mass sources to significantly alleviate the observed
paucity of dwarf galaxies.

Early on after their discovery, the possible extragalactic nature of HI High 
Velocity Clouds (HVCs) was considered (see previous work reviewed by Wakker 
\& van Woerden 1997). More recently, the
idea was forcefully revived by two groups: Blitz \etal ~(1999) 
and Braun \& Burton (1999). The latter postulated that the so--called Compact 
HVCs (CHVCs), a subset of the HVC population defined by a relatively small 
angular extent and apparent spatial isolation are the 
baryonic counterparts of isolated, low mass halos distributed across the LG. 
Hereafter we shall refer to this idea as the ``minihalo/HVC hypothesis''. 
Two important problems mire this idea. The first is that, if present in 
other nearby groups of galaxies, CHVCs could have been detected 
by extant HI surveys; they have not (e.g. Pisano \etal
2007). The second is that the structural properties of the CHVCs
are inconsistent with those expected for low mass halos, according to the 
$\Lambda$CDM structure formation scenario, as pointed out by Sternberg, 
McKee \& Wolfire (2002, hereafter SMW02):
as minihalos, CHVCs appear underconcentrated and, at typical LG distances 
--- on order of 1 Mpc --- they would be too large. 

A widely adopted explanation of the discrepancy between 
the numbers of observed and predicted dwarf galaxies is that gas accretion 
onto low mass halos is suppressed by the intergalactic UV radiation field 
(Ikeuchi 1986; Rees 1986; cf. Hoeft \etal 2006). In their 
{\it Mare Nostrum} simulations, the latter
show that the baryon to total mass fraction may fall steeply 
from the cosmic value of $\simeq 0.16$ to only a few percent between halo 
masses $10^{10}$ \msun ~and $M<10^9$ \msun. SMW02 showed that most of the 
gas in low mass halos 
should be found in a thermally stable, ionized phase which envelops a warm
but neutral and stable component of much lower mass. Within the latter, a
cold core may be able to form and partially convert into stars. A $10^9$ \msun 
~halo may thus contain fewer than $10^6$ \msun ~in a phase detectable
optically or via its 21 cm emission, thus explaining the paucity of such
objects in wide field surveys. The simulations by Ricotti (2009) improve
somewhat the observational prospects: he points out that, as the Universe 
expands, the intergalactic medium cools and the UV radiaton background 
is diluted; changes in the IGM Jeans mass and the halo mass concentration then may
make it possible for a previously `dormant' low mass halo to resume accreting 
gas at low $z$. He proposes that the recently discovered, nearby dwarf galaxy 
Leo T (Irwin \etal 2007) may be such an object. At a distance 
of 420 kpc, Leo T is a star forming galaxy with an HI mass of $2.8\times 10^5$ 
\msun, an HI radius of 300 pc, an indicative dynamical mass within the HI 
radius of $\sim 3.3\times 10^6$ \msun, a total mass to V-band luminosity 
within the HI radius of $56$ and a stellar mass of $\sim 1.2\times 10^5$
\msun ~(Ryan--Weber \etal 2008; see also Grcevich \& Putman 2009).  

The precipitous drop in the cold baryon fraction for halos of mass $\lesssim 10^{10}$
\msun ~proposed by the simulations by Hoeft \etal and the models of SMW02 
can explain the shallow slopes of the low end of optical and HI luminosity
functions. It also indicates that the minihalo/HVC hypothesis could only apply
to sources of more extreme properties than previously thought: HI masses of 
$\lesssim 10^6$ \msun, sizes of $\lesssim 1$ kpc and linewidths of $\lesssim 30$ 
\kms. The requirements for detection of such sources are very challenging,
exceeding the combination of sensitivity and spectral resolution of most large--scale
extragalactic HI surveys. However, the currently ongoing Arecibo Legacy extragalactic 
HI survey (ALFALFA) offers an opportunity for detection
of systems matching the requirements mentioned above, albeit only to distances 
of a few Mpc. The question motivating this paper is whether candidates for minihalos 
with a detectable HI component exist. If they do, the safest place to look for them is 
away from a massive galaxy like the MW. We found them, but we still treat the minihalo 
interpretation with skepticism until the detection is achieved of (i) a stellar counterpart or (ii) 
of an internal velocity field that reveals more reliably a dynamical mass than a plain 
linewidth or (iii) of analogous objects in other galaxy groups.

\section {Observations}\label{obs}

We report the results of 21cm HI observations, made with
the Arecibo 305m radio telescope as part of the ALFALFA survey. They cover 
the region between $9.5^h<RA<16.5^h$, $+4^\circ <Dec<+16^\circ$, of 
$\sim$1300 square degrees, with a spectral resolution of $\sim 5.5$ \kms, 
an angular resolution of $\sim 3'.5$ and a sensitivity such that 
an HI mass of $\simeq 5\times 10^4$ \msun ~can be detected at 1 Mpc distance.
A full description of the observational mode of ALFALFA is given in Giovanelli 
\etal (2007), while the definition and goals of the survey are described in 
Giovanelli \etal ~(2005). Aspects of the processing pipeline are described
in Saintonge (2007) and Kent (2008). 

The chosen region of the sky is a strip extending
between $b=35^\circ$ and $b=80^\circ$ in the North Galactic polar cap, well
suited to minimizing the kinematical intrusion of Galactic HI.  
Within that region, we searched for sources not 
associated with known galaxies with the following additional criteria: (i) 
heliocentric velocity $V_\odot \geq 120$ \kms or $V_\odot \leq -110$ \kms 
~(the $V_\odot$ range between -110 and +120 \kms ~was excluded in order to 
avoid confusion with Galactic emission); (ii) major HI diameter $\leq 15'$; 
(iii) no visible low surface brightness connection to features in the -110 to +120 \kms ~range. 
Twentyone ultra--compact clouds (hereafter UCHVCs) with positive velocities  
(6 are unresolved) and 6 clouds with negative velocities were found. All sources 
detected by ALFALFA with 
a signal--to--noise $S/N<10$ were re---observed and confirmed. Figure 1 shows
an integrated flux map and a line profile of one of the detections. A catalog with a detailed 
description of individual sources is presented in Giovanelli \etal (2009, in preparation).  
Table 1 displays the mean and range values for the properties of the 27 clouds, 
scaled by the unknown distance in Mpc, namely: the range in 
heliocentric velocities for positive and negative velocity clouds, in \kms;
the full linewidth at half power $W$, of generally Gaussian shape; the mean radius at the isophote 
encircling 50\% of the flux $R_{HI}$ in kpc; the $\log_{10}$ of the HI mass 
and of the total mass within $R_{HI}$, in solar units; the crossing time 
$\tau_{cross}=2R_{HI}/W$ in Gyr. On the assumption that the clouds are parts 
of self--gravitating systems, the total mass is estimated via
\be
M_{tot}(<R_{HI}) \simeq R_{HI} \sigma^2/G \label{eq:mdyn}
\ee
with  $\sigma=W/2\sqrt{2\ln 2}$. 
The ``minimum intrusion" approach (Giovanelli \etal 2005) adopted for data taking by 
ALFALFA delivers exceptional bandpass stability and allows reliably recovery of flux 
and size estimation for the UCHVCs. The possibility of extended envelopes of diffuse 
gas below the column density limit of the survey cannot however be excluded.

\begin{deluxetable}{lcc} 
\tablewidth{0pt}
\tabletypesize{\scriptsize}
\tablecaption{Cloud Properties \label{tab:parms}}
\tablehead{
\colhead{Property}& \colhead{Avg.}  & \colhead{Range}  
}
\startdata
$V_\odot$, positive, \kms 	&		& 124:320 \\
$V_\odot$, negative, \kms 	&		& -118:-142 \\
$W$                        	& 24 \kms  	& 16:56 \\
$R_{HI} d_{Mpc}^{-1}$    	& 0.75 kpc 	& <0.4:1.6 \\
$\log~[d_{Mpc}^{-2}M_{HI}/M_\odot]$ 	& 5.25  & 4.75:5.97 \\
$\log~\bar N_{HI}[cm^{-2}]$         	& 19.10 & 18.80:19.58 \\
$\log~[d_{Mpc}^{-1}M_{dyn}(<R)/M_\odot]$ & 7.35 & 6.83:7:77 \\
$\tau_{cross} d_{Mpc}^{-1}$	& 0.08 Gyr 	& 0.02:0.32 \\
\enddata
\end{deluxetable}

\begin{figure}
\includegraphics[height=3.1in]{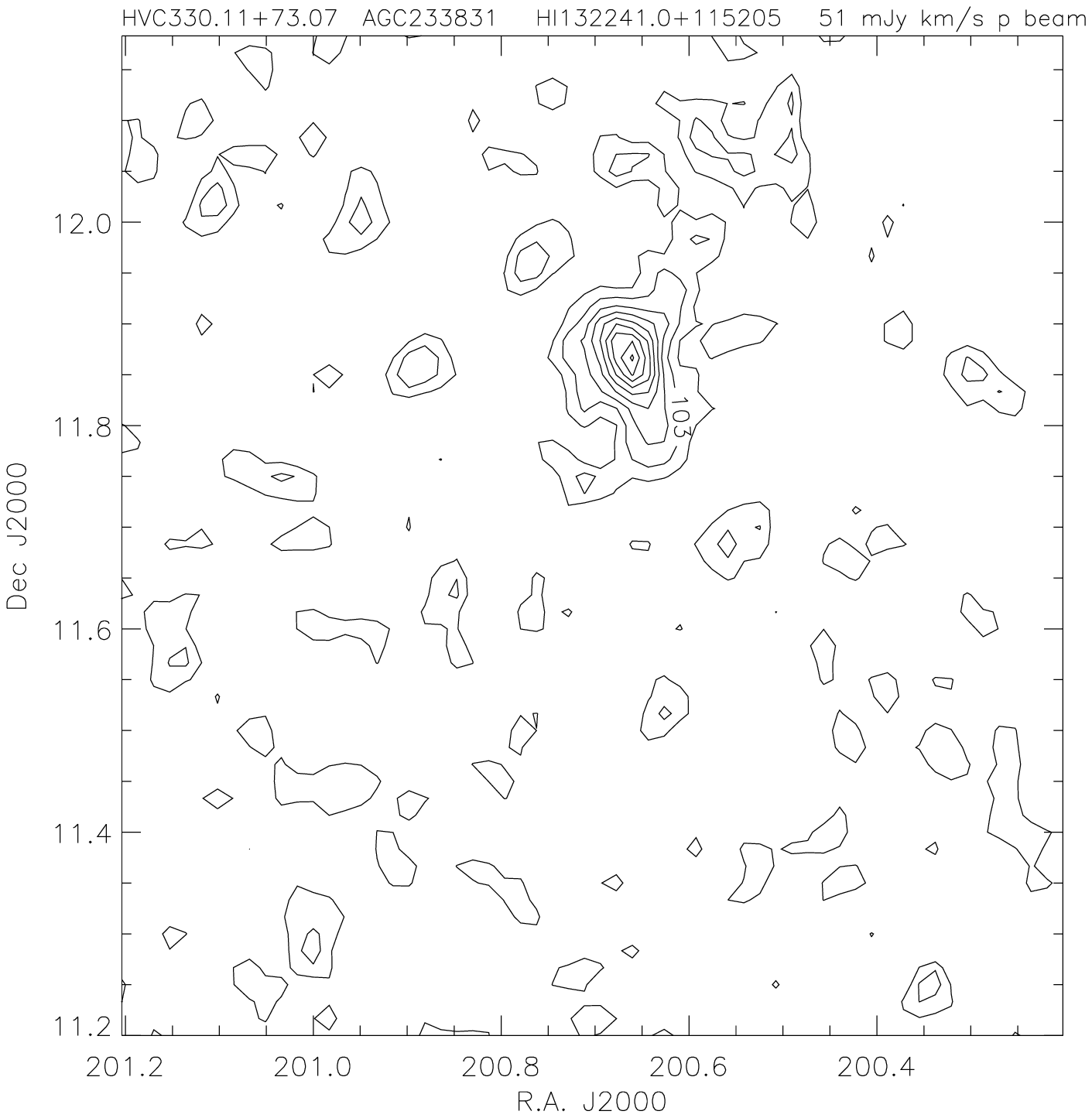}
\includegraphics[height=1.8in,width=3.1in]{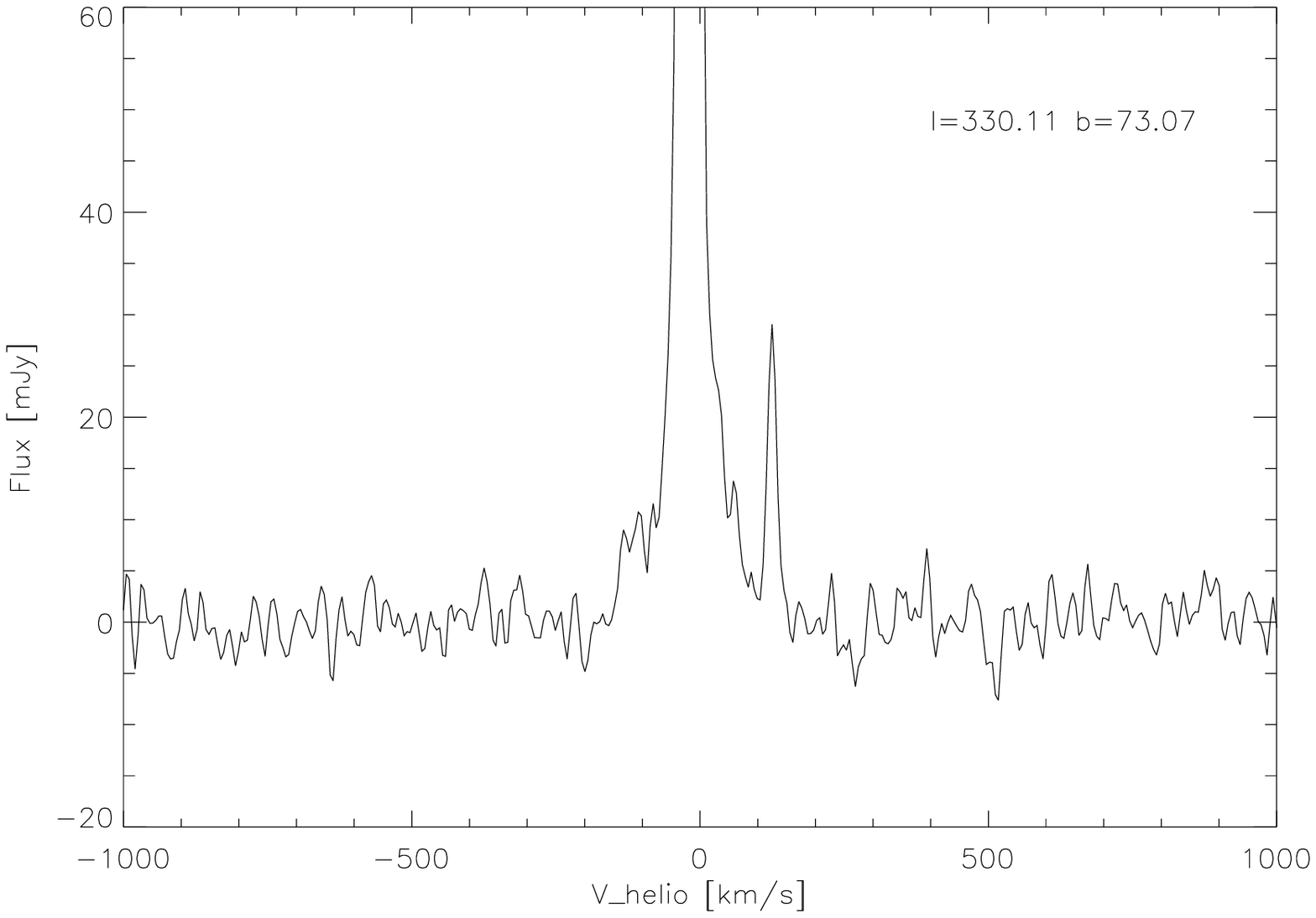}
\caption{{\it Upper panel}: Cloud at $l=330.11$, $b=+73.07$, $cz_\odot=124$ \kms, $W=19$ \kms, angular size of $6'\times 4'$,
integrated flux of 0.60 Jy \kms, $M_{HI}=1.4\times 10^5 d_{Mpc}^2$ \msun, 
$M_{dyn}[<R_{HI}]\simeq 1.1\times 10^7 d_{Mpc}$ \msun. {\it Lower panel}: HI line spectrum of the cloud
shown above. The strong feature at zero velocity is the MW emission;
the cloud is the narrow feature at $cz_\odot=124$ \kms.
}
\label{f1}
\end{figure}

The sky distribution of the UCHVCs is shown in Fig. 2. Also plotted 
are the locations of galaxies with primary distances of $\leq 2.6$ Mpc, as listed in the 
{\it Catalog of Nearby Galaxies} of Karachentsev \etal (2004). Those galaxies are, 
respectively from West to East: D634-03, 
Leo T, Sex B, Leo I, GR8, KKH86 and DDO 187. All but 
Leo I (the nearest at $d=0.25$ Mpc) are detected in HI. Leo T and DDO 187 are just outside the Dec. range of the HI
clouds. Figure 3 shows the location of clouds and nearby galaxies
in the Galactic longitude vs. velocity in the Galactic Standard of Rest plane, where 
$V_{gsr}=V_{lsr}+225\sin l\cos b$ and for $V_{lsr}$ we assume a solar motion of 20 \kms towards 
$l=57^\circ$, $b=25^\circ$. The three negative velocity clouds near $RA=16^h$ could possibly be
associated with the extended, perigalactic HVC complexes A and M. 

\begin{figure}
\includegraphics[height=2.0in]{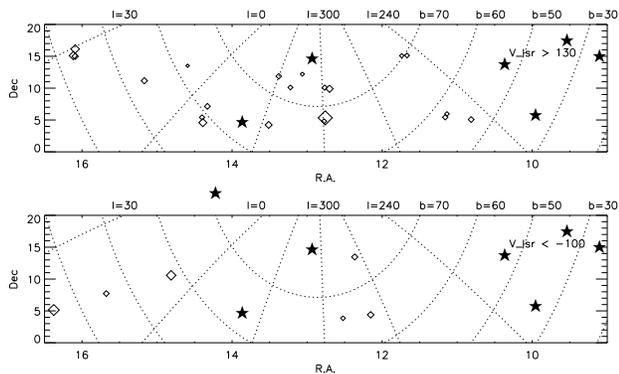}
\caption{Sky distribution of the UCHVCs (small diamonds, of area proportional
to source flux. Galaxies with $d\leq 2.6$ Mpc are plotted as  filled stars. A grid of Galactic
coordinates is superimposed. The star symbol outside the lower box is DDO 187.
}
\label{f2}
\end{figure}

\begin{figure}
\includegraphics[height=2.0in]{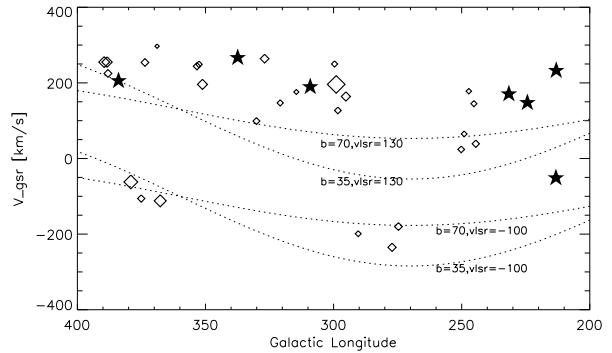}
\caption{Distribution of the UCHVCs in the Galactic longitude vs. Galactic
Standard of rest plane, with superimposed lines of constant LSR velocities, 130 and -100 \kms, 
at two Galactic latitudes, $b=35^\circ$ and $b=70^\circ$. Symbol code as in Figure 2.
}
\label{f3}
\end{figure}

\section {Discussion}\label{discu}

The sky and velocity distributions of the  
clouds found by ALFALFA matches well that of dwarf galaxies in the LG.
If placed at $d=1$ Mpc, the UCHVCs are compact and do not violate 
any structure predictions of the $\Lambda$CDM scenario emphasized by SMW02. 
Their HI masses, $\leq 10^6$ 
\msun, would make them undetectable by previous HI surveys of nearby groups, 
such as that of Pisano \etal (2007). Thus the minihalo/HVC hypothesis could 
apply without posing the observational requirement of LG uniqueness.
HI surveys of nearby groups reaching a sensitivity level of $10^5$ \msun
--- about an order of magnitude more sensitive than ALFALFA --- would be 
necessary to verify the possible existence of similar systems elsewhere than the LG.
Such observations are currently challenging but possible with foreseen upgrades
to current instrumentation.

The mean parameters of the UCHVCs at $d=1$ Mpc are a good match for SMW02
minihalo models with a Burkert density profile, $R_{HI}\simeq 0.7$ kpc,
$M_{HI}\simeq 3\times 10^5$ \msun, total to neutral gas mass ratio of 15, peak 
$N_{HI}\simeq 4\times 10^{19}$ cm$^{-2}$, total halo mass 
$M_{vir}\simeq 3\times 10^8$ \msun, surrounded by a hot, ionized
IGM of presure $P_{HIM}=10$ cm$^{-3}$ K,. 
The peak $N_{HI}$ predicted by the models 
is higher than, but not incompatible with the observed values shown in Table 1, 
as the latter are averaged over the radius of the
clouds and smeared by the $\sim 3.5'$ beam of the Arecibo telescope. The $M_{dyn}$
inferred from the observations are about one order of magnitude smaller 
than the model's $M_{vir}$; yet the two are consistent with each other since the values
inferred from the observations are estimates of the masses within $R_{HI}$.
As the cold baryons dissipatively collapse to the bottom of the halo 
potential well, the extent of the dark matter halo exceeds that of the 
cold gas by a factor of several, explaining the mass discrepancy.

A further test of the minihalo/HVC hypothesis relates to the number of minihalo 
candidates expected from simulations. The cumulative halo mass function $N(>M)$,
is a power law of approximate slope $M^{-0.86}$ (e.g. Fig. 6 of Gottloeber \etal ~2003). 
Figure 4 of Hoeft \etal (2006) shows the transition in the baryon fraction
of halos dropping from the cosmic value of 0.16 to less than 0.02 over a decade
in halo mass, around the ``characteristic mass'' $M_{char}\simeq 10^{9.8}$ \msun, for
which halos are able to retain half their baryons. If we assume that the UCHVCs 
pertain to the category of halos of mass between $M_{char}$ and $M_{char}/\delta_M$, 
and that the $N_{opt}$ optical galaxies in the field are hosted by halos with $M>M_{char}$,
then we can estimate the ratio between the numbers of the two kinds of objects as 
$f_n=N_{opt}/N_{cloud}\simeq \delta_M^{0.86}$, with $\delta_M>1$. The range in dynamical masses
estimated for UCHVCs is about one order of magnitude (see 
Table 1); assuming that the corresponding, putative halo masses are spread just 
as narrowly, we guess $\delta_M\simeq 10$. There are 5 galaxies with $d\leq 2.6$ Mpc
in the region in which the clouds were observed, and the exclusion of the spectral
region $120<V_\odot>-110$ \kms ~in our search due to confusion with Galactic emission
blanks about 50\% of our search volume. The expected number of minihalos
is then $\sim 5\times 10^{0.86}/2 \sim 18$. We identified 27 clouds.
Given the crudeness of the calculation, our cloud detection rate appears
compatible with a minihalo/HVC scenario. Next we ask: what else could the clouds be?

In a Galactic fountain (Shapiro \& Field 1976), gas is accelerated by strong stellar winds and 
supernova explosions in the disk and ejected to the Galactic halo to $z$--heights 
of a few kpc; gas clouds then cool and ``rain back'' onto the disk.
At a distance of 3 kpc from the Galactic plane, the clouds' HI masses 
would be near solar, their sizes on order of 1 or few pc and crossing times 
less than 1 Myr, much shorter than the ballistic timescales of 50--100 Myr. 
Our cloud sample is extracted fron the Galactic polar region, the component 
of the clouds' velocities perpendicular to the Galactic plane is dominant.
Velocities in excess of 200 \kms are difficult to accommodate within a
galactic fountain scenario with currently assumed values for the density 
of a galactic corona (e.g Fukugita \& Peebles 2006; Grcevich \& Putman 2009): 
small clouds would rapidly decelerate due to ram pressure 
and MW tidal forces would eventually disrupt them.

Our survey region overlaps with the 
so--called ``field of streams'' (Belokurov \etal 2006), containing part of the tidally
disrupted remnants of dwarf spheroidal MW satellites, namely the Sag and Orphan
streams. With velocities between +20 and -100 \kms, their kinematics   
are very different from those of most of the UCHVCs, and an
association is unlikely.

The Magellanic Stream is witness of the tidal disruption of the 
Magellanic Clouds by the MW. Most of the Stream is antipodal to the
north Galactic polar region. However, its so-called ``Leading Arm'' (LA)
extends to the northern Galactic hemisphere up to $b\simeq 35^\circ$,
with velocities reaching +250 \kms. If the UCHVCs are a northern extension of 
the LA, according to the tidal model of Connors \etal (2006; see also 
McClure-Griffiths 2008) they would be at distances of order of 100 kpc, have
masses on order of $10^3$ \msun ~and crossing times of $\sim 10$ Myr.
Tidal and ram pressure forces would be mild. This is a plausible scenario,
although the LA would then extend twice as far and forward of the Magellanic
Clouds than previous observations indicated.

Galaxies grow via mergers and intergalactic gas infall. Infalling
gas can be shock--heated to the virial temperature of the halo or fall at colder
temperatures. Simulations show that the cold accretion
mode tends to be more important for small mass halos: for a galaxy like
the MW, $>90$\% of gas accretion is thought to take place in the hot mode
(Keres \etal 2009; Dekel \etal 2009). Sancisi \etal (2008) discuss the
evidence for extraplanar HI in several nearby galaxies and
interpret it as resulting from a combination of mergers of gas rich satellites
and cold gas accretion. In their HI maps, the extraplanar gas is seen 
at distances of 15 kpc or less from the  main galaxy. Two difficulties 
arise with a model whereby our clouds are a manifestation of a similar
phenomenon. First, the objection raised in the discussion of the galactic
fountain scenario, that the clouds are too small, vulnerable and short--lived,
holds in this case as well. Second, as shown by Maller \& Bullock (2004),
thermal instability and conduction prevent the cooling of infalling gas on
scales smaller than the so--called ``Field length'', which for the galactic
corona translates to a mass of $\sim 10^6$ \msun. At perigalactic distances 
of $\sim 100$ kpc or less, our clouds would be much smaller than that limit.

In conclusion, we report the discovery of a category of HVCs which are 
plausible minihalo candidates. They have properties and are found in numbers 
which are compatible with theoretical expectations for halos with masses 
$\lesssim 10^9$ \msun, and could not have been detected by extant HI surveys
beyond the LG. However, it is not yet possible to exclude 
that they may be part of the wider scenario of the yet relatively poorly 
understood perigalactic HVC phenomenon. While difficulties arise in their
interpretations with several more frequently invoked models, reasonable
adjustments could be made to fit some of them, most notably the possibility
that they are an extension of the LA of the Magellanic Stream.
We have thus discovered a category of objects consistent with the minihalo/HVC
hypothesis, but we have not proved that such interpretation is unique.

This work has been supported by the NSF grant AST-0607007
and by a Brinson Foundation grant.

\end{document}